# Relationship between the parent charge transfer gap and maximum transition temperature in cuprates

Wei Ruan[1], Cheng Hu[2], Jianfa Zhao[2], Peng Cai[1], Yingying Peng[2], Cun Ye[1], Runze Yu[2], Xintong Li[1], Zhenqi Hao[1], Changqing Jin[2,4], Xingjiang Zhou[2,4], Zheng-Yu Weng[3,4], Yayu Wang[1,4†]

[1]*State Key Laboratory of Low Dimensional Quantum Physics, Department of Physics, Tsinghua University, Beijing 100084, P. R. China*

[2]*Beijing National Laboratory for Condensed Matter Physics, Institute of Physics, Chinese Academy of Sciences, Beijing 100190, P. R. China.*

[3]*Institute for Advanced Study, Tsinghua University, Beijing 100084, P. R. China.*

[4]*Collaborative Innovation Center of Quantum Matter, Beijing, P. R. China.*

[†] Email: yayuwang@tsinghua.edu.cn

**Abstract:** One of the biggest puzzles concerning the cuprate high temperature superconductors is what determines the maximum transition temperature ($T_{c,max}$), which varies from less than 30 K to above 130 K in different compounds. Despite this dramatic variation, a robust trend is that within each family, the double-layer compound always has higher $T_{c,max}$ than the single-layer counterpart. Here we use scanning tunneling microscopy to investigate the electronic structure of four cuprate parent compounds belonging to two different families. We find that within each family, the double layer compound has a much smaller charge transfer gap size ($\Delta_{CT}$), indicating a clear anticorrelation between $\Delta_{CT}$ and $T_{c,max}$. These results suggest that the charge transfer gap plays a key role in the superconducting physics of cuprates, which shed important new light on the high $T_c$ mechanism from doped Mott insulator perspective.

**Keywords:** cuprates, Mott insulator, charge transfer gap, maximum transition temperature,

scanning tunneling microscopy

**Introduction**

Since the discovery of copper oxide (cuprate) high temperature superconductors about 30 years ago, understanding its microscopic origin has become one of the greatest challenges in condensed matter physics. The problem is well-defined, i.e., to find the mechanism of Cooper pairing for charge carriers residing in the two-dimensional $CuO_2$ plane that is common to all cuprates. However, after decades of experimental and theoretical investigations, very few consensuses have been reached [1-5]. Figure 1A displays the schematic phase diagram of hole-doped cuprates, in which only two well-defined phases are unambiguously determined. First, the parent compound is an antiferromagnetic (AF) Mott insulator due to strong onsite Coulomb repulsion, and the long-range AF order is suppressed by hole doping. Second, the superconducting (SC) phase exists in a range of hole concentrations, and the $T_{c,max}$ is reached at optimal doping near the center of the dome. The definition and doping dependence of other phases, such as the pseudogap phase [6] and charge orders in the underdoped regime [5], are still under much debate.

One of the central issues about the cuprates is what determines the $T_{c,max}$ of a specific system, which varies dramatically from 25 K to 138 K in ambient pressure [7]. A highly robust, but yet mysterious trend is that within each family the double-layer compound has much higher $T_{c,max}$ than the single-layer counterpart. Finding the physical parameter that controls $T_{c,max}$ is apparently a key step for solving the high $T_c$ superconductivity puzzle. For conventional BCS (Bardeen-Cooper-Schrieffer) superconductors, for example, a crucial breakthrough for the final resolution of phonon-mediated mechanism is the discovery of the isotope effect $T_c \sim m^{-1/2}$, where $m$ is the isotopic mass [8, 9]. For the cuprates, there were previous attempts to find the scaling relation between $T_c$ and other physical parameters, such as superfluid density [10] and metallic conductivity near $T_c$ [11]. However, these are macroscopic properties of the SC phase itself, which dictate the phase transition phenomenology but provide little hint regarding the origin of Cooper pairing in the first place. It is highly desirable to find a direct connection between $T_{c,max}$ and the microscopic electronic structure, especially that of the parent compound from where superconductivity emerges.

The most relevant bands of the cuprates derive from the Cu $3d_{x^2-y^2}$ orbital and two O 2$p$ orbitals in the CuO$_2$ plane [12], as schematically drawn in Fig. 1B. Due to strong onsite Coulomb repulsion and relatively small charge transfer energy, the half-filled parent compound is a Mott insulator of the charge transfer type [13]. The hopping integral between O and Cu sites, as well as the onsite and off-site Coulomb repulsions, may all affect the behavior of doped charges hence $T_{c,max}$. Despite this complexity, we notice that the most significant energy scale in the electronic structure of parent cuprate is the charge transfer gap (CTG) between the upper Hubbard band (UHB) and the charge transfer band (CTB, or the Zhang-Rice singlet band), which characterizes the energy needed for an O site electron to migrate to the neighboring Cu sites [14]. If the doped Mott insulator picture is indeed the valid theory for high $T_c$ superconductivity, there might exist a direct link between the CTG size $\Delta_{CT}$ in the parent compound and $T_{c,max}$ at the optimal doping. In a recent theoretical work, the charge transfer energy $\varepsilon_d$ - $\varepsilon_p$ of various cuprates was calculated by first-principles calculations and cluster dynamical mean-field theory [15], and was shown to anticorrelate with $T_{c,max}$. However, so far the relationship between these two important quantities has not been investigated experimentally.

In this paper we use scanning tunneling microscopy (STM) to investigate the CTG in four parent cuprates belonging to the Ca$_{n+1}$Cu$_n$O$_{2n}$Cl$_2$ and Bi$_2$Sr$_2$Ca$_{n-1}$Cu$_n$O$_{2n+4}$ families respectively. We found that within each family, the double-layer compound ($n$ = 2) has much smaller $\Delta_{CT}$ than that of the single-layer ($n$ = 1) compound. The anticorrelation between $\Delta_{CT}$ and $T_{c,max}$ within each family suggests that the Mottness in parent cuprate plays a crucial role in determining the SC properties upon charge doping. In particular, reducing the electron correlation strength to the intermediate regime enhances superconductivity, which is consistent with the pairing mechanism based on the doped Mott insulator picture.

**Methods**

For CCOC single crystals, polycrystalline samples are firstly synthesized by heating the powder mixture of CaO and CuCl$_2$ with a molar ratio of 2:1 at 1073 K for 24 h with intermediate grindings. Then they are heated to 1203 K at a ramp rate of 60 K/h and are kept there for 10 h. Finally, CCOC single crystals are obtained by cooling at a rate of 60 K/h down to room temperature.

For 2-layer CCOC single crystals, the CaO, CuCl$_2$ and CuO powders are firstly mixed with a molar ratio of 3:1:1 and heated to 1223 K at a rate of 180 K/h and kept there for 24 h. The polycrystalline samples are obtained by cooling down to room temperature at a rate of 240 K/h. The 2-layer CCOC single crystals are then obtained with a similar approach as the CCOC, with the reaction temperature raised to 1373 K.

High quality Bi-2201 and Bi-2212 single crystals are grown by the traveling solvent floating zone method [16, 17]. The typical size of the sample is about $2 \times 2 \times 0.1$ mm$^3$, which can be obtained by cleaving the as-grown bulks. Various dopings can be obtained by changing the chemical substitution and post-annealing under different temperatures and atmospheres. For Bi-2212 samples, we obtain nearly insulating samples by annealing in vacuum. All the samples are wrapped with Au foil during the annealing and quenched in liquid N$_2$ immediately after the annealing.

The details of the STM experiments were described elsewhere [18, 19]. All the STS results reported here are obtained at liquid nitrogen temperature $T = 77$ K so that there is finite tunneling current flowing through the highly insulating cuprate crystals. The samples are precooled to 77 K and cleaved in the ultrahigh vacuum chamber with pressure ~ 10$^{-11}$ mbar, and then transferred into the STM stage immediately. An electrochemically etched tungsten tip is calibrated on a clean Au (111) surface before approaching the sample surface. The differential conductance (d$I$/d$V$) curves are obtained by using standard ac lock-in technique with a modulation frequency $f = 423$ Hz.

**Experimental results**

Figure 1C and 1D display the layered crystal structure of the four cuprates studied here. The single-layer oxychloride Ca$_2$CuO$_2$Cl$_2$ (CCOC) is a well-known Mott insulator, and the $T_{c,max}$ can reach 26 K by Na substation of Ca. Its double-layer counterpart is Ca$_3$Cu$_2$O$_4$Cl$_2$ (dubbed "2-layer CCOC" hereafter), in which the two CuO$_2$ planes are separated by a Ca layer, and Na doping leads to a $T_{c,max}$ ~ 49 K [20]. Bi$_2$(Sr,La)$_2$CuO$_{6+\delta}$ (Bi-2201) and

Bi$_2$Sr$_2$(Ca,Dy)Cu$_2$O$_{8+\delta}$ (Bi-2212) are the single-layer and double-layer compounds of the widely-studied Bi family cuprates with $T_{c,max}$ around 38 K and 95 K, respectively. There are two reasons why we choose to study these four cuprates. First, they can be cleaved easily between two van der Waals bonded neighboring layers, as indicated by the blue planes in Fig. 1C and 1D, resulting in ideal surfaces for STM studies. Second, for these systems high quality single crystals with doping at or close to the parent Mott insulator phase are available. For CCOC and 2-layer CCOC, pristine parent compounds can be grown by the flux method, and the former one has been investigated extensively as a prototypical Mott insulator [18, 21, 22]. For Bi-2201, La replacement of Sr introduces electron doping that reduces the hole density to the highly insulating regime close to the parent phase [16, 23]. For Bi-2212, Dy substitution of Ca and further annealing in vacuum can also make it close to the parent Mott insulator limit [17, 24].

Shown in Fig. 2A inset is the atomically resolved topographic image of CCOC taken with bias voltage $V = -3$ V and tunneling current $I = 8$ pA, in which the exposed Cl atoms form a square lattice. On the CCOC surface, the electron density of states (DOS) is highly uniform in space, as reported previously [18]. Figure 2A displays a spatially-averaged differential conductance (d$I$/d$V$) curve measured on a defect-free area of the CCOC surface. The most prominent feature of the spectrum is a well-defined energy gap around the Fermi level ($E_F$), which is the CTG of this parent Mott insulator. To give a quantitative analysis of the gap size, we take the logarithm of d$I$/d$V$ signal as shown in Fig. 2B, which is a common practice for extracting insulating energy gap from d$I$/d$V$ spectroscopy [25, 26]. The two gap edges show linear behavior in the logarithmic plot, which gives a more accurate fit for the exponentially decayed gap edges. By using the intersection of the fitted dashed blue lines illustrated in Fig. 2B, the CTG size of CCOC is determined to be $\Delta_{CT} = 2.0$ eV. This value is also in quantitative agreement with that of a single CuO$_2$ layer grown on Bi-2212 substrate by molecular beam epitaxy technique [27].

We then performed the same measurements on the 2-layer CCOC Mott insulator. The topographic image (taken with $V = -3$ V and $I = 1$ pA ) in Fig. 2C inset also shows a square lattice of the surface Cl layer. Figure 2C displays a spatially-averaged d$I$/d$V$ curve measured on a defect-free area of 2-layer CCOC surface. The CTG size is extracted to be $\Delta_{CT}$ = 1.4 eV, significantly smaller than that of single-layer CCOC. In Fig. 2D we put the d$I$/d$V$ curves of the two samples together, which reveal striking similarities between them. They both show a well-defined energy gap with steep gap edges. The $E_F$ lies close to the occupied states, or the CTB, and is much further away from the unoccupied UHB. The negative half (occupied states) of the two curves are quite similar, but the positive half shows a more pronounced shift.

Figure 3A and 3B insets are the atomically resolved topographic images of Bi-2201 and Bi-2212 scanned using bias voltages $V$ = -1.2 V and -1.8 V, respectively. The exposed BiO planes of both samples show stripe-like structural super-modulations. In addition to the surface Bi-site vacancies, there are also bright and dark patches in both samples due to the chemical and electronic inhomogeneities that are ubiquitous in underdoped cuprates [28, 29]. Because of the spatial inhomogeneities in these samples, the d$I$/d$V$ curves exhibit more complicated behavior. On the Bi-2201 surface (Fig. 3A), the red d$I$/d$V$ curve is taken on the spot marked by the red dot in the inset, which shows the CTG feature similar to those in CCOC and 2-layer CCOC. The blue curve is taken on the spot marked by the blue dot in the inset, which shows a broad electronic state within the CTG induced by local hole doping. The d$I$/d$V$ curves on the Bi-2212 surface (Fig. 3B) are very similar to those of Bi-2201, with some locations showing the CTG feature (red curve) and some others showing the broad in-gap states (blue curve). The evolution of the in-gap states and their spatial variations are highly interesting in their own rights [30], but are not the focus of the current work. Plotting the CTG-type d$I$/d$V$ curves of the two samples together (Fig. 3C), we find that the CTB changes little but the UHB shows a pronounced shift, in much the same way as that in Fig. 2D. The CTG size also decreases significantly from $\Delta_{CT}$ = 1.5 eV in Bi-2201 to $\Delta_{CT}$ = 1.0 eV in Bi-2212.

The results shown above clearly demonstrate that within each cuprate family, the double-layer compound has much smaller CTG size than their single-layer counterpart. In Fig. 4 we summarize the $\Delta_{CT}$ value of the four cuprates studied here as a function of their respective $T_{c,max}$. There is an apparent anticorrelation between the two quantities within each family: the smaller the $\Delta_{CT}$ in the parent Mott insulator, the higher the $T_{c,max}$ at the optimal doping. The overall trend is consistent with the theoretically predicted anticorrelation between the charge transfer energy $\varepsilon_d - \varepsilon_p$ and $T_{c,max}$ [15].

**Discussion and conclusions**

The experimental trend observed here implies that the $T_{c,max}$ of a cuprate at optimal doping is related to the CTG size in its parent compound. This is very surprising because the latter is a much bigger characteristic energy as compared to the former, and the two regions are well-separated in the phase diagram. We propose that the anticorrelation between $\Delta_{CT}$ and $T_{c,max}$ actually sheds important new lights on the mechanism of superconductivity in the cuprates. From the doped Mott insulator perspective, the superexchange coupling $J$ between local moments is responsible for the spin singlet pairing [31]. In the three-band model, $J$ has a complicated expression as a function of the bare charge transfer energy $\varepsilon_d - \varepsilon_p$ and bare Hubbard $U$ [1, 14]. Our observation suggest the CTG size is the most significant single parameter that determines the SC properties upon doping. In the scenario of single-band Hubbard model (Fig. 1B), we may define an effective superexchange $J_{eff} \sim 4t_{eff}^2/\Delta_{CT}$, where $t_{eff}$ is an effective hopping term characterizing the charge transfer process. Therefore the smaller the $\Delta_{CT}$, the larger the $J_{eff}$ and hence the stronger the pairing strength. In Fig. 4 inset we plot $J_{eff} \sim 1/\Delta_{CT}$ as a function of $T_{c,max}$, which has an approximately linear behavior (here we ignore the variation of $t_{eff}$ for different cuprates). From the doped Mott insulator perspective, the CTG is necessary because it protects the local moments by putting a large energy penalty on double occupancy of electrons, which is the origin of strong correlation. However, if the CTG size is too large, it will weaken the pairing strength hence lowering the $T_{c,max}$ that can be achieved. We note that in Bi-2212, the highest $T_c$ cuprate studied here, the CTG size is merely 1.0 eV, putting it in the intermediately correlated regime.

The next question is what factors determine the CTG size in the parent compound, which varies dramatically in the four samples studied here. Because the structure of a single $CuO_2$ plane is nearly the same in all cuprates, it is quite obvious that the out-of-plane environment is mainly responsible for the CTG size variations. As has been shown by first-principles calculations, a key factor is the negatively charged apical ions [32]. Unlike the single-layer compound, which has two apical ions locating symmetrically outside the $CuO_2$ plane, double-layer compound only has one apical ion outside each $CuO_2$ plane, thus renders smaller $\Delta_{CT}$. This scenario explains the experimental observation that within each family, double-layer compound always has much smaller $\Delta_{CT}$ than the single-layer one. Another experimental trend is that the CCOC compound always has a larger $\Delta_{CT}$ than the Bi-family with the same number of $CuO_2$ plane. The main difference between the two families is the change of apical ion from Cl to O, as well as the distance between the apical ion and the $CuO_2$ plane. It is not straightforward why the former situation leads to a larger $\Delta_{CT}$ than the latter, but apparently it is consistent with the first-principles calculations based on the realistic crystal structures [15].

The main message revealed by the observed relationship between $\Delta_{CT}$ and $T_{c,max}$ is that the fate of a particular cuprate family is encoded in its parent compound. This conclusion highlights the importance of Mottness in cuprate high temperature superconductors. Our experiments indicate that the parent CTG size is the most relevant energy scale for the origin of superconductivity. More importantly, $\Delta_{CT}$ can be measured accurately by STM experiment, thus providing a stringent constraint on available theories. Moreover, the sensitivity of $\Delta_{CT}$ to the out-of-plane environment suggests possible route to control $\Delta_{CT}$, and consequently $T_{c,max}$ [33-35]. Strategy for enhancing superconductivity along this direction has been proposed theoretically [32], and is awaiting experimental confirmation.

**Acknowledgement**

This work is supported by NSFC and MOST of China. X. Z. thanks financial support from the Strategic Priority Research Program (B) of the CAS (Grant No. XDB07020300).

**References**:

[1]  P. A. Lee, N. Nagaosa, X.-G. Wen (2006) Doping a Mott insulator: Physics of high-temperature superconductivity. Rev. Mod. Phys. **78**:17-85

[2]  P. W. Anderson, in *Princeton Series in Physics* (Princeton University Press, Princeton, New Jersey 08540, 1997).

[3]  A. Damascelli, Z. Hussain, Z.-X. Shen (2003) Angle-resolved photoemission studies of the cuprate superconductors. Rev. Mod. Phys. **75**:473-541

[4]  N. P. Armitage, P. Fournier, R. L. Greene (2010) Progress and perspectives on electron-doped cuprates. Rev. Mod. Phys. **82**:2421-2487

[5]  E. Fradkin, S. A. Kivelson, J. M. Tranquada (2015) Colloquium: Theory of intertwined orders in high temperature superconductors. Rev. Mod. Phys. **87**:457-482

[6]  T. Timusk, B. Statt (1999) The pseudogap in high-temperature superconductors: an experimental survey. Rep. Prog. Phys. **62**:61-122

[7]  H. Eisaki, N. Kaneko, D. L. Feng et al (2004) Effect of chemical inhomogeneity in bismuth-based copper oxide superconductors. Phys. Rev. B **69**:064512

[8]  C. A. Reynolds, B. Serin, W. H. Wright et al (1950) Superconductivity of Isotopes of Mercury. Phys. Rev. **78**:487-487

[9]  E. Maxwell (1950) Isotope Effect in the Superconductivity of Mercury. Phys. Rev. **78**:477-477

[10] Y. J. Uemura, G. M. Luke, B. J. Sternlieb et al (1989) Universal correlations between $T_c$ and $n_s/m^*$ (carrier density over effective mass) in high-$T_c$ cuprate superconductors. Phys. Rev. Lett. **62**:2317-2320

[11] C. C. Homes, S. V. Dordevic, M. Strongin et al (2004) A universal scaling relation in high-temperature superconductors. Nature **430**:539-541

[12] V. J. Emery (1987) Theory of high-$T_c$ superconductivity in oxides. Phys. Rev. Lett. **58**:2794-2797

[13] J. Zaanen, G. A. Sawatzky, J. W. Allen (1985) Band gaps and electronic structure of transition-metal compounds. Phys. Rev. Lett. **55**:418-421

[14] F. C. Zhang, T. M. Rice (1988) Effective Hamiltonian for the superconducting Cu oxides. Phys. Rev. B **37**:3759-3761

[15] C. Weber, C. Yee, K. Haule et al (2012) Scaling of the transition temperature of hole-doped cuprate superconductors with the charge-transfer energy. Europhys. Lett. **100**:37001

[16] J. Meng, G. Liu, W. Zhang et al (2009) Growth, characterization and physical properties of high-quality large single crystals of $Bi_2(Sr_{2-x}La_x)CuO_{6+\delta}$ high-temperature superconductors. Supercond. Sci. Technol. **22**:045010

[17] X. F. Sun, S. Ono, X. Zhao et al (2008) Doping dependence of phonon and quasiparticle heat transport of pure and Dy-doped $Bi_2Sr_2CaCu_2O_{8+\delta}$ single crystals. Phys. Rev. B **77**:094515

[18] C. Ye, P. Cai, R. Yu et al (2013) Visualizing the atomic-scale electronic structure of the $Ca_2CuO_2Cl_2$ Mott insulator. Nat. Commun. **4**:1365

[19] W. Ruan, P. Tang, A. Fang et al (2015) Structural phase transition and electronic structure evolution in $Ir_{1-x}Pt_xTe_2$ studied by scanning tunneling microscopy. Science Bulletin **60**:798-805

[20] Y. Zenitani, K. Inari, S. Sahoda et al (1995) Superconductivity in $(Ca, Na)_2CaCu_2O_4Cl_2$, The new simplest double-layer cuprate with apical chlorine. Physica C **248**:167-170

[21] J. J. M. Pothuizen, R. Eder, N. T. Hien et al (1997) Single Hole Dynamics in the $CuO_2$ Plane at Half Filling. Phys. Rev. Lett. **78**:717-720

[22] F. Ronning, C. Kim, D. L. Feng et al (1998) Photoemission Evidence for a Remnant Fermi Surface and a *d*-Wave-Like Dispersion in Insulating $Ca_2CuO_2Cl_2$. Science **282**:2067-2072

[23] Y. Peng, J. Meng, D. Mou et al (2013) Disappearance of nodal gap across the insulator-superconductor transition in a copper-oxide superconductor. Nat. Commun. **4**:2459

[24] Y. Zhang, C. Hu, Y. Hu et al (2016) In situ carrier tuning in high temperature superconductor $Bi_2Sr_2CaCu_2O_{8+\delta}$ by potassium deposition. Science Bulletin **61**:1037-1043

[25] R. M. Feenstra (1994) Tunneling spectroscopy of the (110) surface of direct-gap III-V semiconductors. Phys. Rev. B **50**:4561-4570

[26] R. M. Feenstra, J. A. Stroscio, edited by W. J. Kaiser and J. A. Stroscio (Academic Press, Inc., San Diego, 1993).

[27] Y. Zhong, Y. Wang, S. Han et al (2016) Nodeless pairing in superconducting copper-oxide monolayer films on $Bi_2Sr_2CaCu_2O_{8+\delta}$. Science Bulletin **61**:1239-1247

[28] S. H. Pan, J. P. O'Neal, R. L. Badzey et al (2001) Microscopic electronic inhomogeneity in the high-$T_c$ superconductor $Bi_2Sr_2CaCu_2O_{8+x}$. Nature **413**:282-285

[29] K. M. Lang, V. Madhavan, J. E. Hoffman et al (2002) Imaging the granular structure of high-$T_c$ superconductivity in underdoped $Bi_2Sr_2CaCu_2O_{8+\delta}$. Nature **415**:412-416

[30] P. Cai, W. Ruan, Y. Peng et al (2015) Visualizing the evolution from the Mott insulator to a charge ordered insulator in lightly doped cuprates. Nature. Phy., DOI: 10.1038/NPHYS3840.

[31] P. W. Anderson (1987) The Resonating Valence Bond State in $La_2CuO_4$ and Superconductivity. Science **235**:1196-1198

[32] C.-H. Yee, G. Kotliar (2014) Tuning the charge-transfer energy in hole-doped cuprates. Phys. Rev. B **89**:094517

[33] I. Yamada, A. A. Belik, M. Azuma et al (2005) Single-layer oxychloride superconductor $Ca_{2-x}CuO_2Cl_2$ with A-site cation deficiency. Phys. Rev. B **72**:224503

[34] C.-Q. Jin, X.-J. Wu, P. Laffez et al (1995) Superconductivity at 80 K in $(Sr,Ca)_3Cu_2O_{4+\delta}Cl_{2-y}$ induced by apical oxygen doping. Nature **375**:301-303

[35] H. Hobou, S. Ishida, K. Fujita et al (2009) Enhancement of the superconducting critical temperature in $Bi_2Sr_2CaCu_2O_{8+\delta}$ by controlling disorder outside $CuO_2$ planes. Phys. Rev. B **79**:064507

**Figure Captions:**

**Fig. 1. Phase diagram of hole-doped cuprates and crystal structure of the four materials studied here.** (A) Schematic electronic phase diagram of hole-doped cuprates, showing the parent Mott insulator, the AF phase and the SC dome. The maximum SC transition temperature is marked by the orange dot. (B) Schematic band structure of parent cuprate. The charge transfer gap $\Delta_{CT}$ is the energy distance between the UHB and the CTB. (C) The crystal structure of $Ca_2CuO_2Cl_2$ (CCOC) and $Ca_3Cu_2O_4Cl_2$ (2-layer CCOC), respectively. (D) The crystal structure of $Bi_2(Sr,La)_2CuO_{6+\delta}$ (Bi-2201) and $Bi_2Sr_2(Ca,Dy)Cu_2O_{8+\delta}$ (Bi-2212), respectively.

**Fig. 2. STM/STS results on pristine CCOC and 2-layer CCOC.** (A) d$I$/d$V$ spectrum measured on the pristine CCOC surface, showing the well-defined charge transfer gap. The inset shows the topographic image where d$I$/d$V$ is taken. (B) The logarithm plot of the d$I$/d$V$ curve of CCOC, showing how the CTG is extracted. The gap edges and the floor are fitted linearly (blue dashed lines), and the value of CTG is taken as the energy difference between the two intersecting points. (C) d$I$/d$V$ spectrum showing similar CTG measured on 2-layer CCOC surface as shown in the inset topographic image. (D) Comparing the CTG in the two materials, the occupied band edge shows little change, while the unoccupied band edge displays more pronounced shift. The gap size shrinks significantly from CCOC to 2-layer CCOC.

**Fig. 3. STM/STS results on Bi-2201 and Bi-2212.** (A) and (B) d$I$/d$V$ spectra taken on Bi-2201 and Bi-2212, respectively. The insulating charge transfer gap (red) and broad in-gap states (blue) features show up at locations marked by the colored dots as shown in the insets, which are atomically resolved topographic images of Bi-2201 and Bi-2212. (C) The insulating d$I$/d$V$ spectra of both Bi-2201 and Bi-2212 plotted in the same figure. The CTG

size of Bi-2212 is significantly smaller than that of Bi-2201.

**Fig. 4. Anticorrelation between the CTG and $T_{c,max}$.** The measured charge transfer gap sizes in the above four materials as a function of their respective $T_{c,max}$. The experimental data demonstrates an anticorrelation between the CTG size and $T_{c,max}$ for each family, which is illustrated by the two dashed lines. The inset shows the effective superexchange $J_{eff} \sim 1/\Delta_{CT}$ plotted as a function of $T_{c,max}$, assuming the same $t_{eff}$ for different materials.

Figure 1

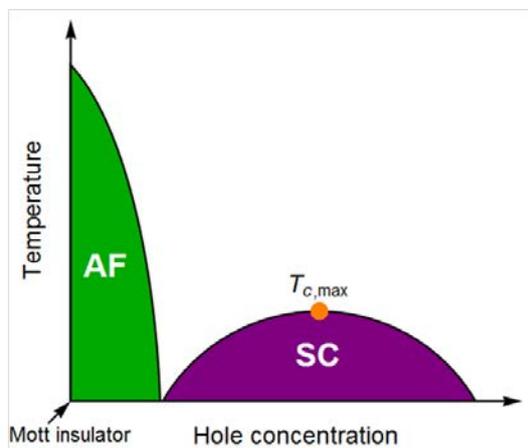

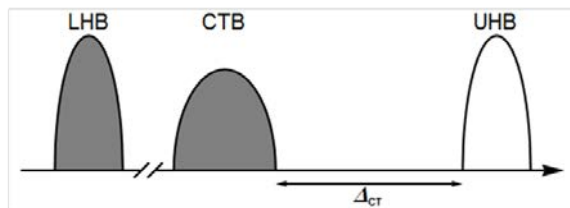

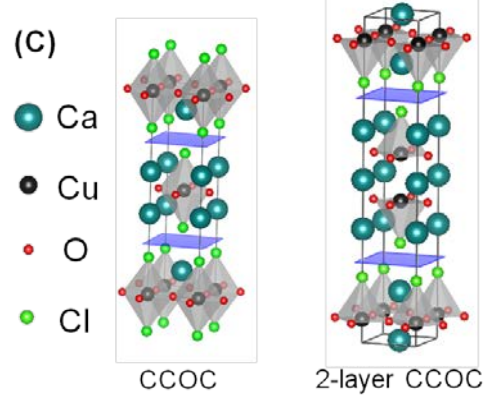

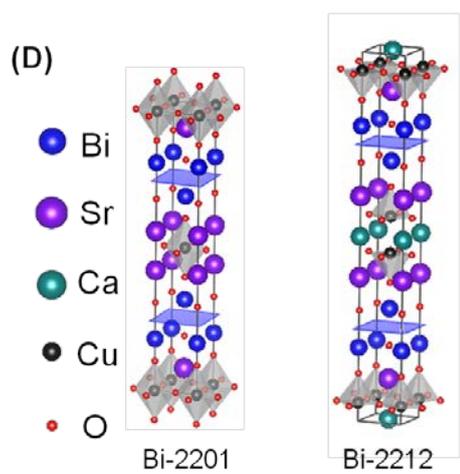

Figure 2

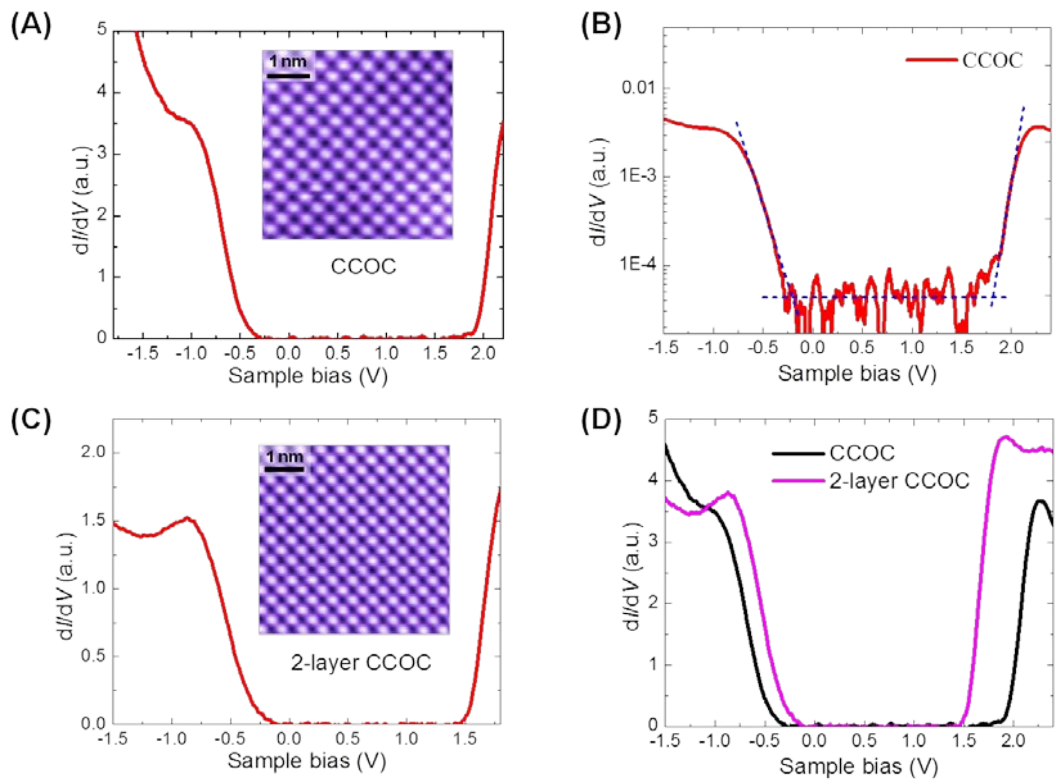

Figure 3

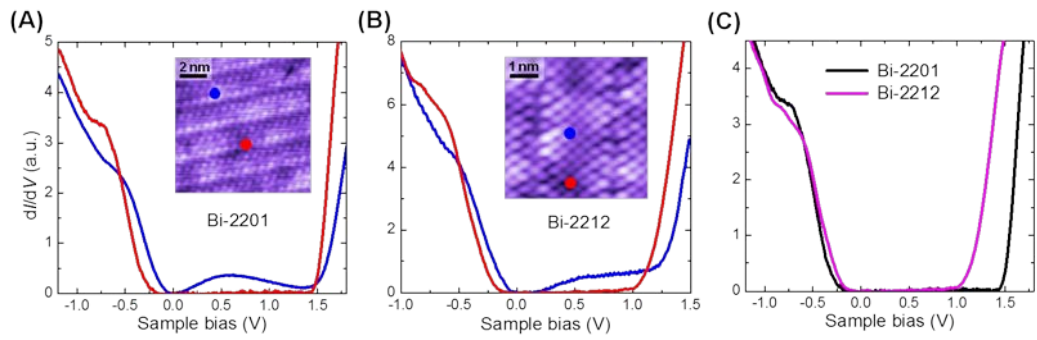

Figure 4

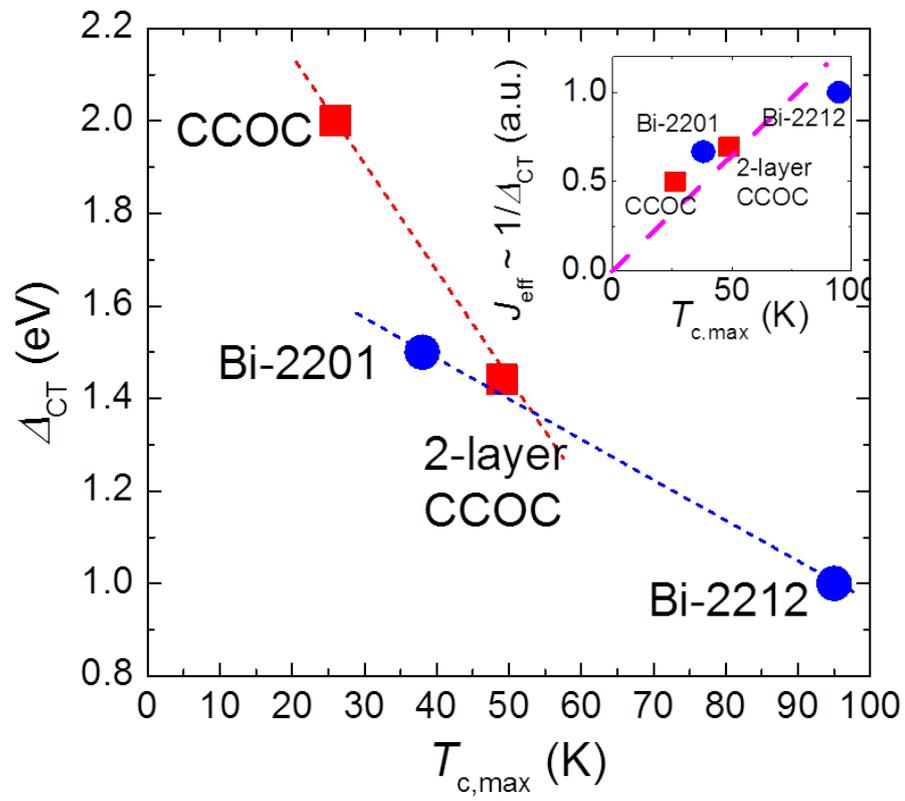